\documentclass[bbm,amsfonts,amsmath,amssymb,12pt]{article}
\usepackage{epsfig}
\usepackage{graphicx}
\usepackage{dcolumn}
\usepackage{bm}
\usepackage{color}
\usepackage{amsmath}
\usepackage{subfigure}
\definecolor{red}{rgb}{1., 0., 0.}
\definecolor{blue}{rgb}{0., 0., 1.}
\definecolor{green}{rgb}{0.1, 0.7, 0.}
\definecolor{purp}{rgb}{1,0,1}

\def\drawbox#1#2{\hrule height#2pt
        \hbox{\vrule width#2pt height#1pt \kern#1pt
              \vrule width#2pt}
              \hrule height#2pt}

\def\Asym#1#2{\vcenter{\vbox{\drawbox{#1}{#2}
              \kern-#2pt       
              \drawbox{#1}{#2}}}}

\newcommand {\beq} {\begin{equation}}
\newcommand {\eeq} {\end{equation}}
 \newcommand{\be}{\begin{eqnarray}}
\newcommand{\ee}{\end{eqnarray}}

\begin{document}

\begin{titlepage}

\begin{flushright}
\small 
NORDITA-2004-73, UPRF-2004-14
\end{flushright}

\vspace {1cm}

\centerline{\Large \bf Information on the Super Yang-Mills Spectrum}

\vskip 1cm \centerline{\large $^a$~A. Feo , $^b$~P. Merlatti and
$^b$~F. Sannino} \vskip 0.1cm

\vskip .5cm \begin{center}$^a$ Dipartimento di Fisica,
Universit\`{a} di Parma and INFN Gruppo \\ Collegato di Parma, Parco
Area delle Scienze, 7/A. 43100 Parma, Italy.\end{center}

 \vskip 0.5cm \centerline{$^b$ NORDITA, Blegdamsvej 17, DK-2100 Copenhagen \O, Denmark }

\vskip 1cm

\begin{abstract}
We investigate the spectrum of the lightest states of $N=1$ Super
Yang-Mills. We first study the spectrum using the recently extended
Veneziano Yankielowicz theory containing also the glueball states
besides the gluinoball ones. Using a simple K\"{a}hler term we show
that within the effective Lagrangian approach one can accommodate
either the possibility in which the glueballs are heavier or lighter
than the gluinoball fields.

We then provide an effective Lagrangian independent argument which
allows, using information about ordinary QCD, to deduce that the
lightest states in super Yang-Mills are the gluinoballs. This helps
constraining the K\"{a}hler term of the effective Lagrangian. Using
this information and the effective Lagrangian we note that there is
a small mixing among the gluinoball and glueball states.


\end{abstract}

\end{titlepage}
\section{Introduction}

Strongly interacting supersymmetric gauge theories are much studied
since, in many respects, they resemble Quantum Chromodynamics.
Theoretically we already know a great deal about supersymmetric
gauge theories which are closer to their non supersymmetric cousins,
namely $N=1$ supersymmetric gauge theories, see
\cite{Intriligator:1995au} for a review. Here we show that it is also
possible to use the present {\it experimental} knowledge of QCD to
learn about super Yang-Mills non perturbative dynamics.

    Gaining information on the spectrum of $N=1$ super Yang-Mills is
    the goal of this paper. We will provide a definite answer to the question of which state is the
    lightest supersymmetric state in super Yang-Mills. We will compare the $R=2$ sector of the
    theory whose representative is the gluinoball superfield with
    the $R=0$ sector, which is a glueball state. We will study the mixing as well.

The tools we will use to make our predictions are: i) The newly
extended Veneziano and Yankielowicz (VY) \cite{Veneziano:1982ah}
effective Lagrangian which, besides the $R=2$ sector, consistently
includes the $R=0$ sector \cite{Merlatti:2004df}; ii) The relation
between QCD with one flavor and super Yang-Mills recently advocated
in \cite{Armoni:2004uu} and further studied in
\cite{Sannino:2003xe}. To make such a correspondence one uses a
large $N$ limit in which a Dirac fermion is in the two index
antisymmetric representation of the $SU(N)$ gauge theory. This limit
was first introduced in \cite{Corrigan:1979xf} and further studied
in \cite{Kiritsis:1989ge}; iii) The knowledge of the QCD spectrum
from experiments and lattice computations.

Our exploration starts with the effective superpotential built in
terms of two chiral superfields constituting the minimal set of
superfields needed to fully describe the vacuum of super Yang-Mills
\cite{Veneziano:1982ah,Merlatti:2004df}. The two chiral superfields
are respectively the $R=2$ gluinoball chiral superfield and the
$R=0$ glueball chiral superfield. To be able to discuss the spectrum
we augment the superpotential by a K\"{a}hler term. We then proceed
to show that within the effective Lagrangian approach one can
accommodate either the possibility in which the glueballs are
heavier or lighter than the gluinoball fields. This is not too
surprising since, even though the superpotential in
\cite{Merlatti:2004df} has no free parameters, the K\"{a}hler term
is not constrained. However once the mass ordering of the states is
determined this effective theory can be used to understand the
mixing between these states.

Using the information about ordinary QCD and the large $N$ relation
discussed above \footnote{Such a correspondence has also been
analyzed within the string theory approach \cite{DiVecchia:2004ev}.
Also recently the phase diagram of theories in higher
representations was carefully studied in \cite{Sannino:2004qp}
providing very interesting new results. In
\cite{Sannino:2004qp,{Hong:2004td}} was also realized the relevance
of strongly interacting theories with fermions in higher
representations for the dynamical breaking of the electroweak
symmetry.} we predict the gluinoball to be lightest state of super
Yang-Mills.

The knowledge of which state is the lightest is then used to
constrain the effective K\"{a}hler term. This allows us to show that
the mixing among the gluinoball and glueball is small.

The paper is organized as follows. {In section 2 we study} the $N=1$
SYM spectrum via the extended VY effective Lagrangian. In section 3
we provide the Lagrangian free prediction of which is the lightest
state of super Yang-Mills while in section 4 we compare our results
with existing lattice results. We finally conclude in section 5.

\section{Spectrum via Extended VY Lagrangian}
Effective Lagrangians are an important tool for describing strongly
interacting theories in terms of their relevant degrees of freedom.
The well known VY effective Lagrangian constructed in
\cite{Veneziano:1982ah} economically describes the vacuum structure
of super Yang-Mills. It concisely summarizes the symmetry of the
underlying theory in terms of a ``minimal'' number of degrees of
freedom which are encoded in the superfield $S$
\begin{eqnarray}
S=\frac{3}{32\pi^2\, N}{\rm Tr}W^2 \ ,
\end{eqnarray}
where $W_{\alpha}$ is the supersymmetric field strength. When
interpreting $S$ as an elementary field it describes a gluinoball
and its associated fermionic partner. In this paper we follow the
notation introduced in \cite{Sannino:2003xe}.

Besides the gluinoballs with non zero $R$-charge also glueball
states with zero $R$ charge are {\it important} degrees of freedom.
These states play a relevant role when breaking supersymmetry by
adding a gluino mass term \cite{Merlatti:2004df}. This is so since
the basic degrees of freedom of the pure Yang-Mills theory are
glueballs. {}Further support for the relevance of such glueball
states in super Yang-Mills comes from lattice simulations
\cite{Feo:2002yi}.  And finally it has been shown in
\cite{Merlatti:2004df} that these new states are needed to provide a
more consistent picture of the vacuum of super Yang Mills via an
effective Lagrangian description.

However no {\it physical} glueballs appear in the VY effective
Lagrangian. In this paper we employ the extended VY Lagrangian,
provided in \cite{Merlatti:2004df}, which takes into account the
glueball states via the introduction of
a chiral superfield $\chi$ with the proper quantum numbers. The
constraint used to construct the effective superpotential involving
$S$ and $\chi$ has been to reproduce the anomalies of super
Yang-Mills while keeping unaltered the vacuum structure. This led to
a general form of the superpotential in terms of an undetermined
function of the chiral field $f(\chi)$. It has, nevertheless, been
possible to motivate a specific form for $f(\chi)$ which has a
number of amusing properties. {}For example the $N$ vacua of the
theory emerge naturally when integrating out the glueball superfield
$\chi$ as suggested first in \cite{Kovner:1997im}. This intriguing
relation seems moreover to have a natural counterpart in the
geometric approach to the effective Lagrangian theory proposed by
Dijkgraaf and Vafa \cite{Dijkgraaf:2002dh}. Another important check
is related to supersymmetry breaking. When adding a gluino mass to
the theory the same choice of the function $f(\chi)$ leads to a
K\"{a}hler independent part of the ``potential'' which has the same
functional form of the glueball effective potential for the non
supersymmetric Yang-Mills theory developed and used in
\cite{schechter}.

In \cite{Merlatti:2004df} the focus was on the properties of the
theory which depends solely on the superpotential which reads:
\begin{eqnarray}\label{ms}
W\left[S,\chi\right] &=&
\frac{2N}{3}S\left[\ln\left(\frac{S}{\Lambda^3}\right)^N -N -N\ln
\left(-e\frac{\chi}{N} \ln\chi^N\right) \right] \ ,
\end{eqnarray}
{ where $S$ is the gaugino bilinear superfield ($S=\varphi +
\sqrt{2} \theta \psi  + \theta^2F$) and $\chi$ describes the $R=0$
glueball type degrees of freedom, ($\chi=\varphi_{\chi} +
\sqrt{2}\theta \psi_{\chi}  + \theta^2 F_{\chi}$)}.


A K\"{a}hler is needed for computing the spectrum. VY suggested the
simplest K\"{a}hler for $S$, i.e. $( SS^{\dag})^{1/3}$ which does
not upset the saturation of the quantum anomalies. We modify the VY
K\"{a}hler to provide a kinetic term also for $\chi$ as follows:
\begin{eqnarray}
K(S,S^{\dag},\chi,\chi^{\dag})~=~
\frac{9\,N^2}{\alpha}(SS^{\dag})^{1/3}(1~+~\gamma
\,\chi\chi^{\dag})\ ,
\end{eqnarray} where $\gamma$ is a positive number. A more general
K\"{a}hler can be constructed with an arbitrary function of $\chi$
and  $\chi^{\dag}$. We however expect the simplest choice to provide
a reasonable description of the mass spectrum.

To compute the potential of
the theory we need the K\"{a}hler metric which is:
\begin{eqnarray}g^{l\bar{m}} = \left[g_{l\bar{m}}^{-1}\right]^T \ ,
\quad {\rm with} \quad g_{l,\bar{m}}=\frac{\partial^2\,K}{\partial
\varphi^l\partial \bar{\varphi}^{\bar{m}}} \ , \end{eqnarray} and
$\varphi^1=\varphi$ while $\varphi^2=\varphi_{\chi}$. The potential
is then
\begin{eqnarray}V\left[\varphi\ ,\varphi_{\chi}\right]= \frac{\partial W}{\partial \varphi^l}\,
g^{l\bar{m}}\, \frac{\partial W^{\dagger}}{\partial
\bar{\varphi}^{\bar{m}}}\ .\end{eqnarray} Given the extended VY
Lagrangian and the simple K\"{a}hler term we introduced we derive
\begin{eqnarray}
V\left[\varphi\ ,\varphi_{\chi}\right]&=& \left(\varphi\bar{
\varphi}\right)^{\frac{2}{3}} \frac{4N^2\,\alpha}{9} \left[\left|
\ln\left(\frac{
\varphi}{-e\varphi_{\chi}\ln\varphi_{\chi}}\right)\right|^2+
\frac{\left(1+\gamma\,\varphi_{\chi}
\bar{\varphi}_{\chi}\right)}{9\gamma} \, \left|\frac{\ln
\varphi_{\chi}
+1}{\varphi_{\chi}\,\ln\varphi_{\chi}}\right|^2\right. \nonumber
\\&& \nonumber \\&&\nonumber \\
&+& \left. \frac{\ln \varphi_{\chi} +1}{3\ln
\varphi_{\chi}}\,\ln\left(\frac{
\bar{\varphi}}{-e\bar{\varphi}_{\chi}\ln\bar{\varphi}_{\chi}}\right)
+ \frac{\ln \bar{\varphi}_{\chi} +1}{3\ln
\bar{\varphi}_{\chi}}\,\ln\left(\frac{
\varphi}{-e\varphi_{\chi}\ln\varphi_{\chi}}\right) \right] \, .
\end{eqnarray}
The potential is bounded from below and has a well defined global
minimum.

We now turn to the spectrum. Since the theory is supersymmetric, it
is sufficient to investigate only the bosonic sector. Holomorphicity
of the superpotential also guarantees degeneracy of the opposite
parity states.

Before providing the mass eigenvalues and eigenstates and to build
up our intuition it is very instructive to consider the two
following limits.

    \subsection{Integrating out $\chi$}
{}Eliminating $\chi$ via its equation of motion at the
superpotential level:
\begin{eqnarray}
\frac{\partial W\left[S,\chi\right]}{\partial \chi} = 0 \, ,
\end{eqnarray}
yields
\begin{eqnarray}
\chi = 1/e \ .
\end{eqnarray}
Here we are restricting ourself to the first branch of the logarithm
in order to determine the spectrum. The effective Lagrangian is the VY one
with $\alpha$ replaced by $\alpha/(1+\gamma/e^2)$. The
supersymmetric spectrum for the $S$ superfield is:
\begin{eqnarray}
M_S= \frac{2}{3} \frac{\alpha}{1+ \gamma/e^2}\Lambda \, .
\end{eqnarray}
\subsection{Integrating out $S$}
It is also interesting to construct the effective Lagrangian for
$\chi$ after having integrated out $S$. 
The equation of motion for $S$ reads:
\begin{eqnarray}
\frac{\partial W\left[S,\chi\right]}{\partial S} = 0
\end{eqnarray}
yielding
\begin{eqnarray}
S=-e\Lambda^3\,\chi\,\ln \chi \,  \ .
\end{eqnarray}
The effective theory for $\chi$ is then
\begin{eqnarray}
{\cal L} &=& \frac{9N^2}{\alpha}\Lambda^2\int d^2 \theta
d^2\bar{\theta} \left[e^2 \chi\chi^{\dagger}\ln \chi \ln
\chi^{\dagger} \right]^{1/3} \left(1+\gamma \chi
\chi^{\dagger}\right)\nonumber \\ &+&\frac{2N^2}{3}e\Lambda^3\int\,
d^2\theta\, \chi \ln \chi + {\rm c.c.} \, ,
\end{eqnarray}
while the common mass of the components of the chiral superfield $\chi$
is
\begin{eqnarray}
M_{\chi} = \frac{2}{27} \frac{\alpha\, e^2}{\gamma} \Lambda \ .
\end{eqnarray}
In the left panel of figure \ref{mass} we plot the spectrum for the
two limiting theories as a function of $\gamma$ after setting
$\alpha$ to one. The dashed line represents $M_S$ after having
integrated $\chi$ out while the dotted line is $M_\chi$ after $S$
has been integrated out. Notice that there is a value of $\gamma$
above which the $\chi$ superfield is lighter than the gluinoball
field $S$.

    \subsection{The spectrum and mixing from the effective theory}
\begin{figure}[h]
\begin{center}
\includegraphics[width=13truecm,height=5truecm,clip=true]{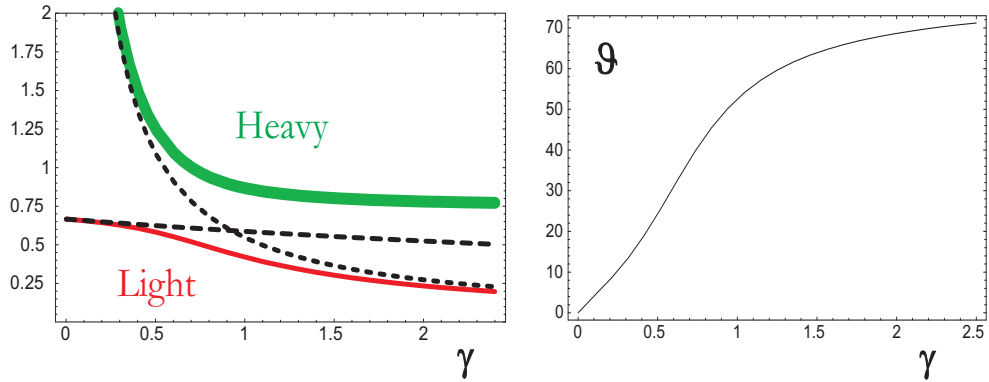}
\caption{Left panel: Physical spectrum as function of $\gamma$. The
dashed ($M_{S}$) and the dotted line ($M_{\chi}$) correspond
respectively to the case in which $\chi$ or $S$ were integrated out.
The two continuous, not straight lines, correspond to the heavy
(thick-line) and light (thin-line) physical eigenvalues. At small
$\gamma$ we have the glueball state heavier than the gluinoball
state. {}For $\gamma$ large enough there is an inversion and the
glueball state is lighter than the gluinoball state. Right Panel:
Mixing angle as function of $\gamma$} \label{mass}
\end{center}
\end{figure}

Diagonalizing the full potential we derived the mass eigenvalues.
They are plotted in the left panel of figure \ref{mass} as a
function of the unknown parameter $\gamma$ while we fixed $\alpha$
to one. { The two continuous lines} correspond to the physical
eigenvalues obtained within our theory. At small $\gamma$ we have
the glueball state heavier than the gluinoball state. It is amusing
to observe how well the limiting case for $\chi$ approximates the
physical spectrum at small and large $\gamma$. Comparing with the
glueball spectrum, obtained after having integrated out the $S$
field, we can say that for $\gamma$ large enough there is an
inversion and the glueball state is lighter than the gluinoball
state. In order to make the statement more precise we define the
physical states as:
\begin{eqnarray}
|{\rm Light}\rangle &=&\cos\vartheta\, |R=2\rangle +
\sin\vartheta\,|R=0\rangle \ ,\nonumber \\
&& \nonumber \\ |{\rm Heavy}\rangle &=&-\sin\vartheta\, |R=2\rangle
+ \cos\vartheta\,|R=0\rangle \ ,
\end{eqnarray}
where $|{\rm Light(Heavy)}\rangle$ corresponds to the lightest
(heaviest) eigenstate. The state with $R=2$ is the pure gluinoball
and the $R=0$ state is the pure glueball type state. The mixing
angle \footnote{To determine the mixing angle we have first
diagonalized the kinetic term. This yields the first contribution to
the mixing between the $R=2$ and $R=0$ states. We have canonically
normalized the resulting states and finally diagonalized the
resulting potential. The angle presented in the figure is the
resulting mixing angle due to the combined action of the two
rotation matrices needed to fully diagonalize the system.} as
function of $\gamma$ is presented in the right panel of figure
\ref{mass}.

At large $\gamma$ the lower curve corresponds mainly to an $R=0$
state. At small $\gamma$ we have that the lightest state is mainly
$R=2$ while the heavy one is an $R=0$ state. We can then say that at
small $\gamma$ the gluinoball is the lightest of the two chiral
superfields.
Although our superpotential has no free parameters the spectrum
depends on the K\"{a}hler coefficient $\gamma$.


\section{Using QCD to determine the lightest super Yang Mills state.}
{}From the previous analysis we learned that the extended VY
superpotential \cite{Merlatti:2004df} while correctly describing the
vacuum of SYM alone is not sufficient to disentangle the puzzle of
which state is the lightest in super Yang-Mills. We have shown that
the poor knowledge of the K\"{a}hler term, when two heavy chiral
superfield are considered simultaneously, heavily affects the
effective Lagrangian ability of making general predictions for the
mass ordering of the chiral supermultiplets.

We note, however, that in the literature
\cite{Farrar:1997fn,{Farrar:1998rm},Cerdeno:2003us} it has been
argued, using an effective Lagrangian constructed via a three form
superfield approach, that the glueball states, in the supersymmetric
limit are lighter than the gluinoball states. This would correspond
to the region of $\gamma > 1$ in our case.

The burning question is: Can we construct an argument in favor of a
certain general ordering pattern which does not rely on the
K\"{a}hler's ambiguities of the effective Lagrangian theory ?

The answer is positive. In fact we can make use of a recent
correspondence which, at large $N$, maps the non barionic and
bosonic sector of {$SU(N)$ Yang Mills theory with one massless Dirac
fermion in the two index antisymmetric (symmetric) representation in
the bosonic sector of SYM} \cite{Armoni:2004uu}. Interestingly the
two index antisymmetric representation for three colors is QCD with
one flavor \cite{Corrigan:1979xf}.
 To be more specific we can map directly the complex gluinoball state into
 the ordinary $\eta^{\prime}$ particle and the associated scalar partner of one flavor
 QCD. The super Yang-Mills glueball states are also mapped directly in the ordinary glueball states.

In \cite{Sannino:2003xe} one was also able to compute the leading
$1/N$ corrections and make more quantitative predictions from the
use of the large $N$ correspondence for fermions in the two index
antisymmetric as well as the two index symmetric representation of
the gauge group.

When restricting our attention to the two index antisymmetric theory
which interpolates between QCD and super Yang-Mills, till now such a
relation has been used to make \cite{Armoni:2004uu,Sannino:2003xe}
statements about QCD {from the knowledge of super Yang-Mills}.

Here we will do the converse, i.e. we use well known properties of
QCD to make physical predictions for SYM. Since this is a large $N$
type correspondence we admit, upfront, that our predictions are
affected by a 30\% error which should however be confronted with the
ignorance of the effective Lagrangian's K\"{a}hler term as well as
the not definitive lattice results. Notice that the poor knowledge
of the K\"{a}hler does not reduce the power of the effective
Lagrangian for SYM. Having a compact description of gluino and
glueball state and the associated vacuum properties is still very
relevant. The approach of this subsection complements the effective
Lagrangian one while providing new constraints on the K\"{a}hler
structure.

We start by considering QCD with one massless flavor. Here the low
lying composite fields made prevalently of glue are heavier than the
low lying mesons while the true lightest states are the
$\eta^{\prime}$ and the associated scalar meson\footnote{There is a
well known mixing between $\eta^{\prime}$ and $\eta$ \cite{PDG}
which can be neglected here.}.

We can be more precise. We can estimate the one flavor
$\eta^{\prime}$ mass by taking the experimental value of the 3
flavor QCD eta prime mass reported on the Particle Data Group (PDG)
which is
$(957.78 \pm 0.14) $ Mev \cite{PDG} and then use the
Veneziano-Witten formula:
\begin{eqnarray}
M^2_{\eta^{\prime}}[\rm 1~flavor] = \frac{1}{3}
M^2_{\eta^{\prime}}[\rm 3~flavor] \ .
\end{eqnarray}
{}For the scalar partner the experimental situation is more delicate
\cite{Black:1998wt}. Although one might be tempted to use the mass
of the particle $f_0(600)$ quoted in the PDG \cite{PDG} and
investigated in much detail in \cite{Sannino:1995ik}, this state is
not a $q\bar{q}$ state \cite{Harada:2003em}. A better candidate for
the scalar partner of the $\eta^{\prime}$ is the $f_0(1370)$
\cite{PDG,{Sannino:1995ik}}, which is heavier than the
$\eta^{\prime}$. This is also consistent with the predictions of
\cite{Sannino:2003xe}. We can average the scalar and pseudoscalar
mass for three flavors and then use again the Veneziano-Witten
formula, yielding as a common mass:
\begin{eqnarray}
M_{\rm \bar{q}q}[\rm 1~flavor] \approx 672.~{\rm MeV} \ .
\end{eqnarray}
{}From the lattice formulation of pure Yang-Mills we have that the
lightest scalar glueball is in the range of $1730 $ Mev
\cite{Morningstar:1999rf} and the
pseudoscalar glueball is $2590 $ Mev \cite{Morningstar:1999rf}
yielding as a common averaged mass
\begin{eqnarray}
M_{\rm glueball} \approx 2.16~{\rm GeV} \ .
\end{eqnarray}
We now neglected the small $N_f$ dependence. It is then clear that
the glueballs are much heavier than the lightest massive scalar and
pseudoscalar $\bar{q}q$ state.

This is even more evident when considering the large N expansion a
la 't Hooft \footnote{We thank R. Narayanan for suggesting also this
argument.}. Here it turns out that the $\eta^{\prime}$ becomes very
light at large N while the glueball masses remain large and do not
scale to zero. We conclude that the splitting between the low lying
mesonic states and the glueball states in one flavor QCD is large in
comparison to the invariant scale of the theory:
\begin{eqnarray}
\Delta_{\rm 1-Flavor} = \frac{M_{\rm glueball} -
M_{\bar{q}q}}{\Lambda} \approx 5-7 \ ,
\end{eqnarray}
for a $\Lambda \approx 200-300$~MeV.


We then expect that in super Yang-Mills the splitting between the
glueball and the gluinoball is
\begin{eqnarray}
\Delta_{\rm SYM} \sim \Delta_{\rm 1-Flavor} \quad {\rm
up~to~30\%~corrections} \ .
\end{eqnarray}

\begin{figure}[h]
    \begin{center}
      \epsfig{file= 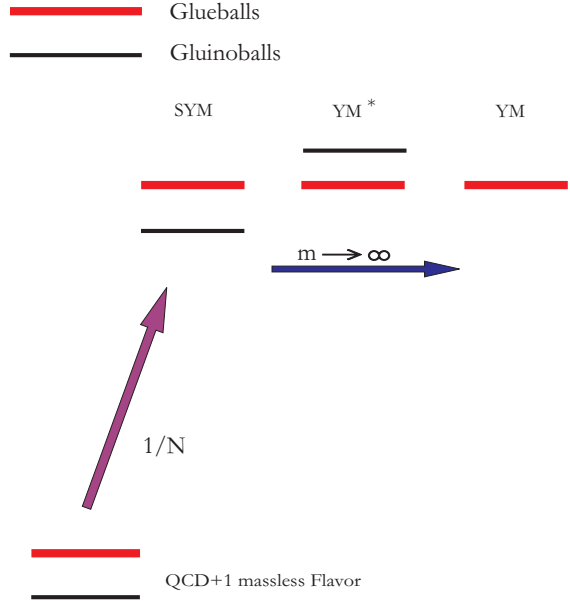, scale=0.39}
      \caption{Spectrum of the theory as a function of the gluino
      mass. The thick (thin) line represents the glueball (gluinoball) mass scale.
      Using the correspondence with QCD (lower end of the diagram) we set the relative
      ordering of the spectrum in the super Yang Mills case. We then add a gluino. There
      will, certainly, be a value of the gluino mass of the order of the
      confining scale above which the gluinoball states is
      heavier than the glueball state. The corresponding theory is
      indicated in the figure with YM$^{\ast}$. {}For sufficiently
      large gluino masses the gluino decouples and one is left
      with the pure YM theory.
     }
      \label{fig-tc}
    \end{center}
\end{figure}

This finally suggests that the lightest state in super Yang Mills
is the gluinoball. When adding a mass to the gluino we
expect that at sufficiently large masses, with respect to the
invariant scale of the theory, the gluinoball states becomes heavier
than the glueball field. In figure \ref{fig-tc} we provide the
spectrum as a function of the gluino mass.

The information that the glueball is heavier than the gluinoball
implies that the coefficient $\gamma$ appearing in our K\"{a}hler is
less than one. This means that the mixing between the $R=0$ sector,
i.e. the glueballs, and the $R=2$ is small.

\section{Spectrum of super Yang-Mills theory from lattice results}
In this section we compare our previous expectations with Monte
Carlo results for $N=1$ super Yang-Mills theory with dynamical light
gluinos,
\cite{Campos:1999du,Farchioni:2001wx,Montvay:2001aj,Peetz:2002sr,FP},
where the bound state mass spectrum is investigated. The formulation
of Curci and Veneziano \cite{Curci:1986sm} which uses Wilson-type
fermions has been used here
\footnote{Simulations of super Yang-Mills with domain wall fermions
were studied in \cite{fkv}. }.

In the numerical simulations with light gluinos a gluino bare mass
is introduced which breaks supersymmetry softly. In the Curci and
Veneziano formulation the supersymmetric limit coincides with the
chiral limit and by studying, for example, the pattern of chiral
symmetry breaking, through the study of the first order phase
transition of the gluino condensate it is then possible to determine
the value of the critical hopping parameter which corresponds to the
supersymmetric (chiral) limit (i.e. zero gluino mass). It is also
possible to determine the gluino massless limit from the study of the
supersymmetric Ward-Takahashi identities.

The restoration of supersymmetry can be checked through the study of
the supersymmetry multiplets. Lattice simulations,
\cite{Campos:1999du,Farchioni:2001wx,Montvay:2001aj,Peetz:2002sr,FP},
have been performed for four scalar degrees of freedom and two
Majorana fermions. They show a non-trivial mass spectrum of the SYM
theory with an $SU(2)$ gauge group. More in detail,
\cite{Montvay:2001aj}, the lightest bound state with almost the same
mass are the glueball $O^+$ and the pseudoscalar component of the
gluinoball field, while the heavier supermultiplet contains the
pseudoscalar glueball state $O^-$ and the scalar gluinoball, also
degenerate in mass. The mass difference between states of opposite
parity is
bigger than the gluino mass used for the simulations
\cite{Farchioni:2001wx}. This implies that it does not seems to be
an effect of softly broken supersymmetry, as has been also stressed
in \cite{Cerdeno:2003us}.


Recently, \cite{FP}, results for larger lattices near the
supersymmetric point are presented. The analysis of the spectrum in
\cite{FP} shows, interestingly, that the pseudo scalar gluinoball is
lighter than the other two particles of the lightest supermultiplet
at the value of the gluino mass measured. The latter findings seem
to be more consistent with our results. These also give further
evidence that the older lattice simulations were describing an
intermediate regime in which supersymmetry is still badly broken.

\section{Conclusions}

We have investigated the spectrum of the lightest states of $N=1$
super Yang-Mills. The spectrum was first studied using the recently
extended Veneziano Yankielowicz theory containing also the glueball
states besides the gluinoball ones. We have shown that by adopting a
simple K\"{a}hler term the effective Lagrangian approach can
accommodate either the possibility in which the glueballs are
heavier or lighter than the gluinoball fields. To resolve the
ambiguity we have provided an effective Lagrangian independent
argument. We used the information about ordinary (one-flavor) QCD and
the recent
map into the bosonic sector of SYM \cite{Armoni:2004uu} to deduce
that the lightest states in super Yang-Mills are, indeed, the
gluinoballs. This observation helps constraining the K\"{a}hler term
of the effective Lagrangian. Using this information and the
effective Lagrangian we have then shown that there is a small mixing
among the gluinoball and glueball states.

Finally we conclude that the lightest state is the gluinoball field
and it has a small mixing with the glueball state. This supports the
use of the VY effective theory with the inclusion of the glueball
state \cite{Merlatti:2004df} needed to provide a more consistent
description of the non perturbative aspects of the super Yang-Mills
vacuum. Due to the small mixing between the glueball and the
gluinoball it is reasonable to compute the super Yang-Mills spectrum
via the effective Lagrangians at zero and non zero gluino masses
\cite{Masiero-Veneziano} or for the orientifold theories at finite
$N$ \cite{Sannino:2003xe}. Our results also indicate that previous
lattice simulations were still far from the supersymmetric limit.

\vskip 1cm \centerline{\bf Acknowledgments} We thank P.H. Damgaard,
L. Del Debbio, P. Di Vecchia, B. Lucini, P. de Forcrand, R.
Narayanan, M. Shifman and M. Teper for useful discussions.

\end{document}